\documentclass{appolb}

\usepackage{amssymb} 
\usepackage{amsmath} 
\usepackage{txfonts}
\usepackage[pdftex]{graphicx}
\usepackage{xspace}
\usepackage{sidecap}

\newcommand{\jpsi}{\ensuremath{\mathrm{J}/\psi}\xspace}
\newcommand{\psiP}{\ensuremath{\psi \mathrm{(2S)}}\xspace}
\newcommand{\snn}{\ensuremath{\sqrt{s_{\mathrm{\scriptscriptstyle NN}}}}\xspace}
\newcommand{\s}{\ensuremath{\sqrt{s}}\xspace}
\newcommand{\pt}{\ensuremath{p_{T}}\xspace}

\newcommand{\cff}[1]{{Fig.~\ref{#1}}}

\begin{document}
\title{The latest STAR results on quarkonium production
\thanks{Presented at Excited QCD 2015}
}
\author{Barbara Trzeciak, for the STAR Collaboration
\address{Faculty of Nuclear Sciences and Physical Engineering,\\Czech Technical University in Prague}
\\
{
}
\address{}
}
\maketitle
\begin{abstract}
We report the latest results of J/$\psi$, $\psi(2S)$ and $\Upsilon$ production in the dielectron decay channel at mid-rapidity, from the STAR experiment. We present \jpsi cross section measurements in $p+p$ collisions at \s = 200 and 500 GeV, as well as the first measurement of the $\psi(2S)$ to \jpsi ratio at \s = 500 GeV.
We also show \jpsi and $\Upsilon$ production in heavy ion collisions: \jpsi nuclear modification factors ($R_{AA}$) in Au+Au collisions at $\sqrt{s_{NN}} =$ 200, 62.4 and 39 GeV and in U+U collisions at $\sqrt{s_{NN}} =$ 193 GeV, $\Upsilon$ $R_{AA}$ in $d+$Au, Au+Au and U+U collisions at $\sqrt{s_{NN}} =$ 200, 200 and 193 GeV, respectively.
The results are compared to different model calculations.
\end{abstract}
  
\section{Introduction}
It was proposed that in high energy heavy-ion collisions quarkonium states can be used as probes of Quark-Gluon Plasma (QGP) formation. Due to the Debye-like color screening of the quark-antiquark potential in the hot and dense medium, quarkonia are expected to dissociate and this ``melting'' can be a signature of the presence of a QGP~\cite{Matsui:1986dk}. 
Studies of production of quarkonium states in heavy-ion collisions can provide insight into the thermodynamic properties of the hot and dense medium~\cite{Mocsy:2007jz}.
However, in the medium created in heavy-ion collisions quarkonium yields can be enhanced, relative to $p+p$ collisions, due to statistical recombination of heavy quark-antiquark pairs. Also, effects related to the ''normal'' nuclear matter, so called cold nuclear matter (CNM) effects: (anti-)shadowing, initial-state parton energy loss, final state nuclear absorption or the Cronin effect, can alter the qarkonium production. Therefore, it is difficult to distinguish the color screening effect from the other effects. 
J/$\psi$ with $p_{T} >$ 5 GeV/$c$, are expected to be almost not affected by the recombination and CNM effects~\cite{Zhao:2010nk} at the RHIC energies. Also $\Upsilon$ are considered as cleaner probes of the QGP, compared to J/$\psi$, because of the negligible statistical recombination and co-mover absorption~\cite{Rapp:2008tf,Adamczyk:2013poh}.
Systematic measurements of quarkonium production as a function of centrality and transverse momentum, for different colliding systems and collision energies may help to understand the quarkonium production mechanisms in heavy-ion collisions as well as properties of the created medium.

\section{Charmonium measurements}
\label{sec:charmoniumMeasurements}

STAR has measured inclusive \jpsi production in $p+p$ collisions at \s = 200 and 500 GeV and \psiP production at \s = 500 GeV through the dielectron decay channel at $|y| <$ 1. The \pt spectrum measurements at \s = 200 GeV are shown on the left panel of \cff{fig:Jpsipp}, where low-~\cite{Kosarzewski:2012zz} and high-\pt~\cite{Adamczyk:2012ey} results are shown separately as squares and circles, respectively. Data points are compared to different models of \jpsi production. 
The predictions of the Color Evaporation Model (CEM)~\cite{Frawley:2008kk} and NLO calculations within the nonrelativistic QCD (NRQCD) effective field theory that include both color-singlet and color-octet contributions (NLO CS+CO)~\cite{Ma:2010jj}, describe the data quite well. Both predictions are for the prompt \jpsi production.
The Color Singlet Model (CSM)~\cite{Artoisenet:2007qm} for the direct production underpredicts the measured cross section. On the other hand, RHIC \jpsi polarization measurements~\cite{Adamczyk:2013vjy} show different trend than the prediction of NRQCD~\cite{Chung:2009xr}, and are in agreement with the CSM~\cite{Lansberg:2010vq}.
The right panel of \cff{fig:Jpsipp} shows \jpsi cross section as a function of $x_{T}$ (defined as $x_{T} = 2p_{T}/\sqrt{s}$) at \s = 500 GeV, for \pt range of 4-20 GeV/$c$. The STAR result (full circles) is compared to measurements at different energies. The \jpsi cross section follows the $x_{T}$ scaling: $\frac{d^{2}\sigma}{2 \pi p_{T} dp_{T} dy} = g(x_{T})/(\sqrt{s})^{n}$, with $n =$ 5.6 $\pm$ 0.2 at mid-rapidity and $p_{T} >$ 5 GeV/$c$ for a wide range of colliding energies~\cite{Abelev:2009qaa}.
The left panel of \cff{fig:JpsiPsi} shows a ratio of $\psi(2S) / J/\psi$ production (full circle) compared to measurements of other experiments at different colliding energies, in $p+p$ and $p+$A collisions. The STAR result is consistent with the observed trend and, with the current precision, no collision energy dependence of the $\psi(2S)$ to J$/\psi$ ratio is observed.

\begin{figure}[ht]
		\includegraphics[width=0.49\linewidth]{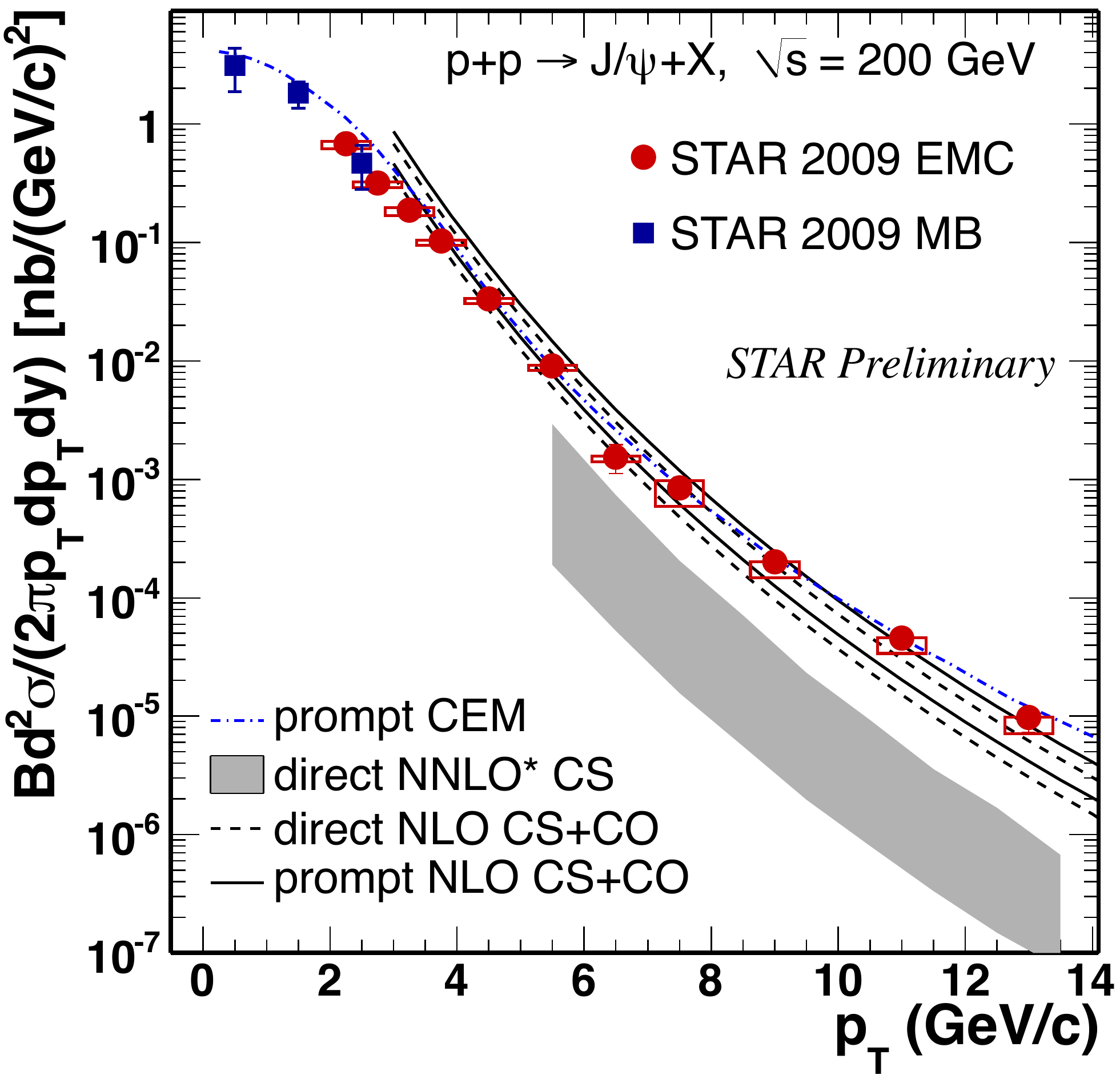}
		\includegraphics[width=0.45\linewidth]{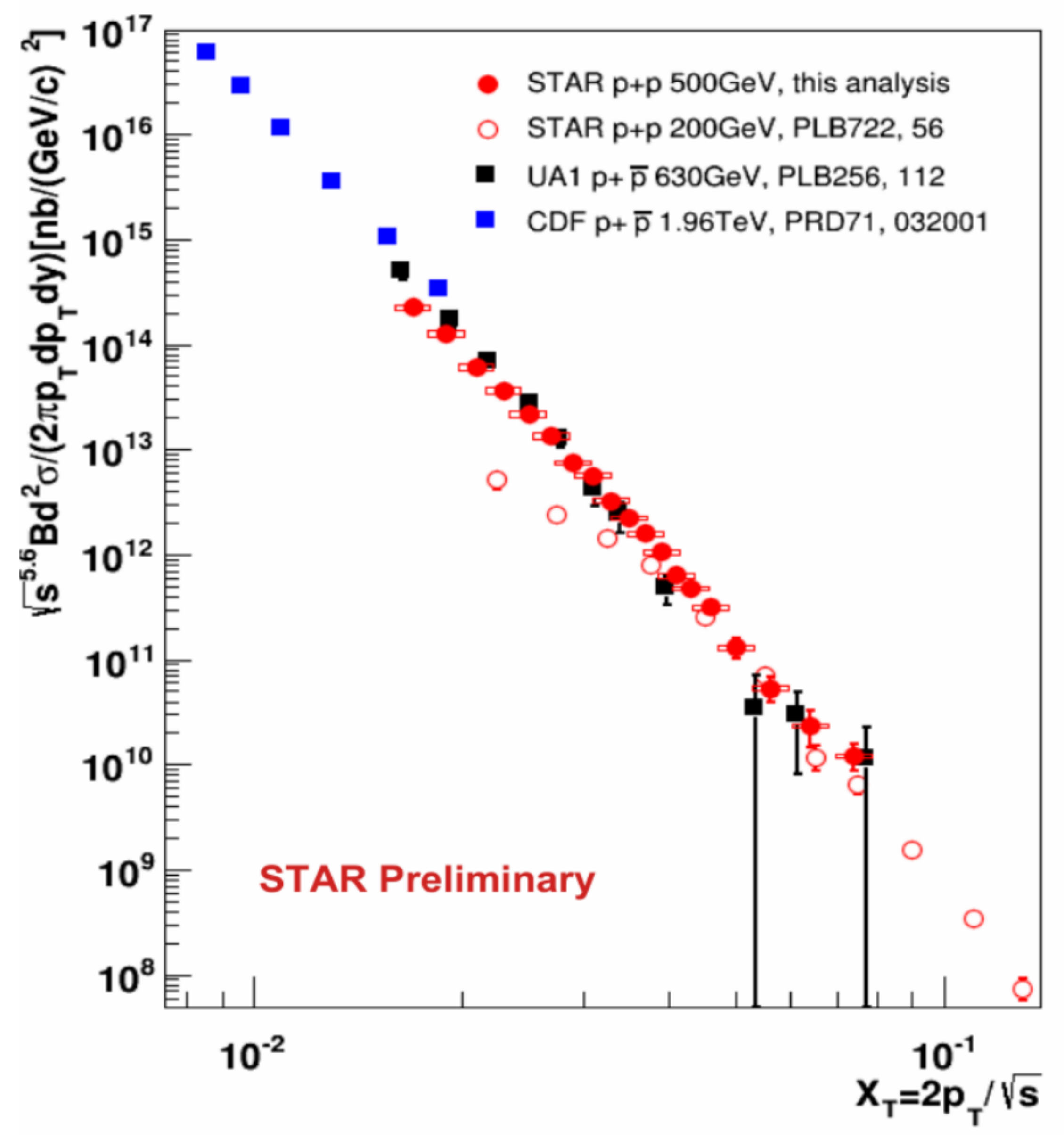}
		\caption{Left: J/$\psi$ invariant cross section vs $p_{T}$ in $p$+$p$ collisions at $\sqrt{s} =$ 200 GeV at mid-rapidity at low~\cite{Kosarzewski:2012zz} and high $p_{T}$~\cite{Adamczyk:2012ey} shown as squares and circles, respectively, compared to different model predictions~\cite{Frawley:2008kk,Ma:2010jj,Artoisenet:2008fc}. Right: J/$\psi$ invariant cross section multiplied by $\sqrt{s}^{5.6}$ vs $x_{T}$ in $p$+$p$ collisions at $\sqrt{s} =$ 500 GeV at mid-rapidity shown as full circles compared to measurements at different energies.
}
		\label{fig:Jpsipp}
\end{figure}

\begin{SCfigure}
		\centering
		\caption{Ratio of $\psi(2S)$ to J/$\psi$ in $p+p$ collisions at $\sqrt{s} =$ 500 GeV from STAR (the full circle) compared to results from other experiments at different energies.}
		\includegraphics[width=0.5\textwidth]{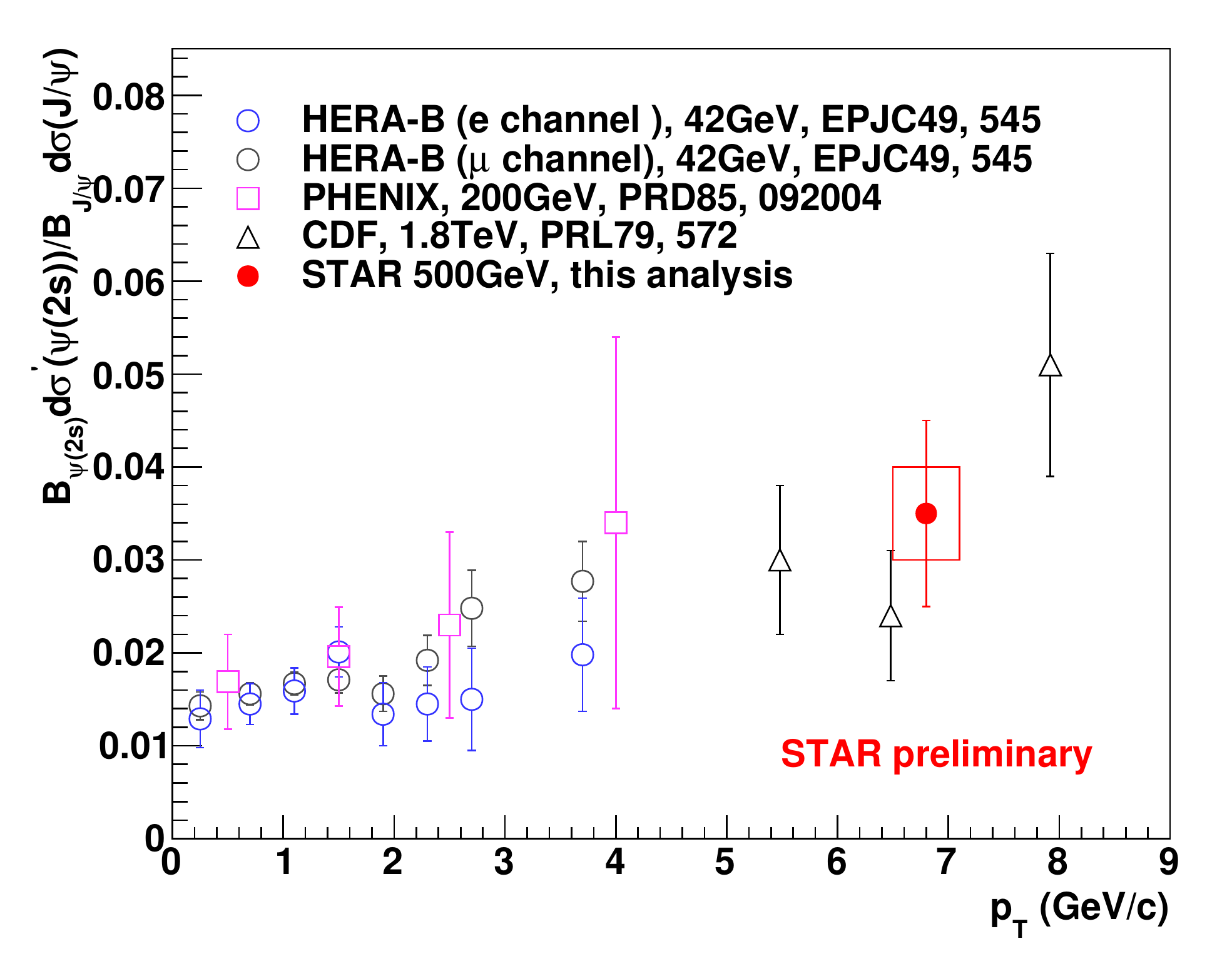}	
		\label{fig:JpsiPsi}
\end{SCfigure}

Effects of the hot and dense medium created in heavy-ion collisions are evaluated using the nuclear modification factors ($R_{AA}$), defined as the ratio of the particle yield in $A+A$ collisions to that in $p+p$ collisions, scaled by the number of binary collisions. \jpsi $R_{AA}$ measurements have been done in Au+Au and U+U collisions, as a function of transverse momentum, and for different centrality bins (represented as the number of participant nucleons, $N_{part}$, in a collision).
First, we show measurements of $R_{AA}$ in Au+Au collisions at \snn = 200 GeV vs $N_{part}$, for low- ($<$ 5 GeV/$c$)~\cite{Adamczyk:2013tvk} and high-$p_{T}$ ($>$ 5 GeV/$c$)~\cite{Adamczyk:2012ey} inclusive \jpsi production. As can be seen on the left panel of \cff{fig:JpsiRaa}, $R_{AA}$ decreases with increasing centrality, and the high-$p_{T}$ J/$\psi$ suppression level is lower that the low-$p_{T}$ one. Since high-$p_{T}$ J/$\psi$ are expected to be almost not affected by the recombination or CNM effects at RHIC energies, the suppression in central collisions points to the color screening effect and thus formation of the QGP.
The Liu {\it{et al.}}~\cite{Liu:2009nb} model is in agreement with the low- and high-\pt $R_{AA}$ results, while the prediction of Zhao and Rapp~\cite{Zhao:2010nk} is consistent with low-\pt data but underpredicts the high-\pt $R_{AA}$. Both models take into account direct J/$\psi$ production with the color screening effect and production via the recombination of $c$ and $\bar{c}$ quarks. 

\begin{figure}[ht]
		\centering
		\includegraphics[width=0.45\textwidth]{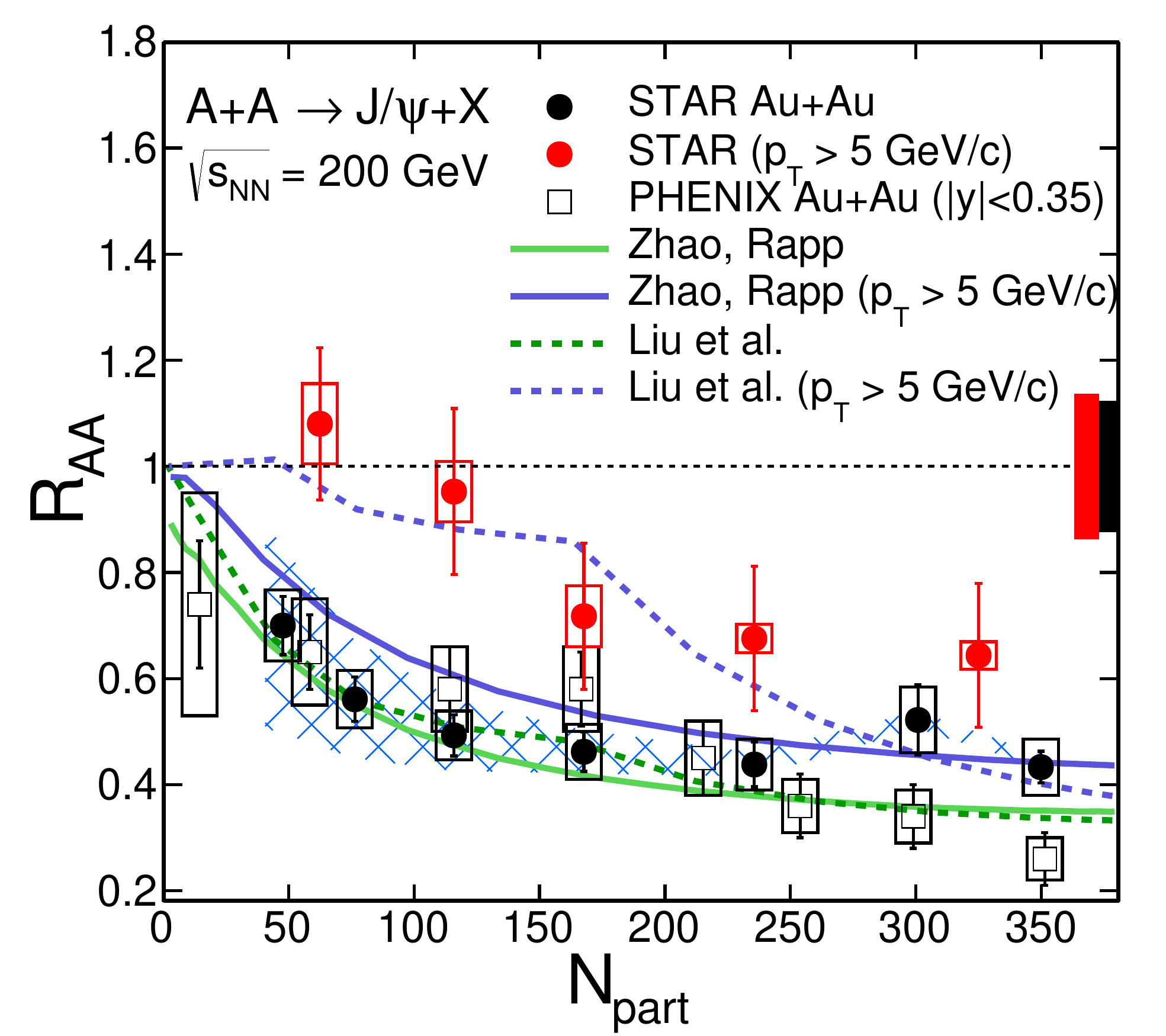}
		\includegraphics[width=0.49\textwidth]{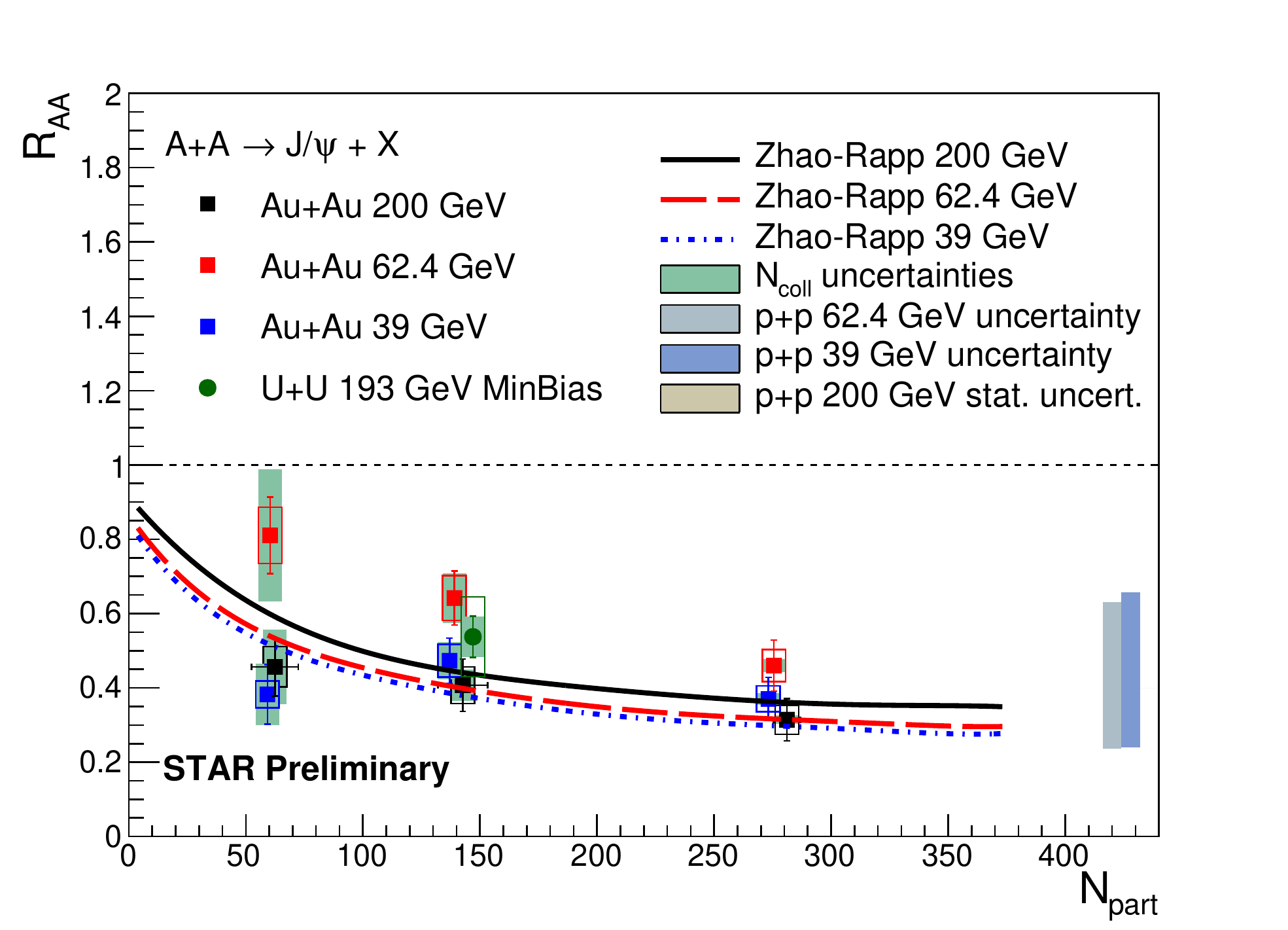}
		\caption{Left: Low- ($<$ 5 GeV/$c$) and high-$p_{T}$ ($>$ 5 GeV/$c$) J/$\psi$ $R_{AA}$ as a function of $N_{part}$ in Au+Au collisions at $\sqrt{s_{NN}} =$ 200 GeV at mid-rapidity (\cite{Adamczyk:2012ey,Adamczyk:2013tvk}) with two model predictions (\cite{Zhao:2010nk,Liu:2009nb}). Right: J/$\psi$ $R_{AA}$ as a function of $N_{part}$ in Au+Au collisions at $\sqrt{s_{NN}} =$ 200, 62.4 and 39 GeV at mid-rapidity with model predictions. As a full circle the minimum bias U+U measurement at $\sqrt{s_{NN}} =$ 193 GeV is also presented.}
		\label{fig:JpsiRaa}
\end{figure}

The interplay between recombination, CNM effects and direct J/$\psi$ production can be studied by changing energies of colliding ions.
The right panel of Fig.~\ref{fig:JpsiRaa} shows low-$p_{T}$ J/$\psi$ $R_{AA}$ in Au+Au collisions for different colliding energies, $\sqrt{s_{NN}} =$ 200, 62.4 and 39 GeV. The observed suppression is similar for all these energies, and results are well described by the model of Zhao and Rapp~\cite{Zhao:2010nk}.
The significant uncertainties come from lack of precise $p+p$ measurements at 62.4 and 39 GeV, instead Color Evaporation Model calculations~\cite{Nelson:2012bc} are used as baselines.
STAR has also performed an analysis of J/$\psi$ $R_{AA}$ in U+U collisions at $\sqrt{s_{NN}} =$ 193 GeV, where the initial energy density can be up to 20\% higher than in Au+Au collisions, in the same centrality bin~\cite{Kikola:2011zz}. The centrality-integrated U+U result is shown in the right panel of Fig.\ref{fig:JpsiRaa}, as a full circle. The suppression level is consistent with what is measured in Au+Au collisions.

\section{$\Upsilon$ measurements in $d+$Au and A+A collisions}
\label{sec:UpsilonMeasurements}

$\Upsilon$ measurements in STAR have been performed in $p+p$, $d$+Au and Au+Au collisions at $\sqrt{s_{NN}} =$ 200 GeV~\cite{Adamczyk:2013poh} and in U+U collisions at $\sqrt{s_{NN}} =$ 193 GeV. 
The left panel of \cff{fig:upsilon} shows the nuclear modification factor for $d+$Au collisions, $R_{dAu}$, as a function of rapidity. The result is compared to CEM calculations with shadowing based on the EPS09 nPDF parametrization~\cite{Frawley:2008kk} presented as the shaded area, model of Arleo {\it{et al.}} where suppression of $\Upsilon$ is due to initial-state parton energy loss~\cite{Arleo:2012rs} presented as the dashed line, and the model combining both shadowing and energy loss, presented as the dashed-dotted line. The strong $\Upsilon$ suppression at $y \sim 0$ cannot be explained by the current predictions of CNM effects in $d+$Au collisions.

\begin{figure}[ht]
		\centering
		\includegraphics[width=0.4\textwidth]{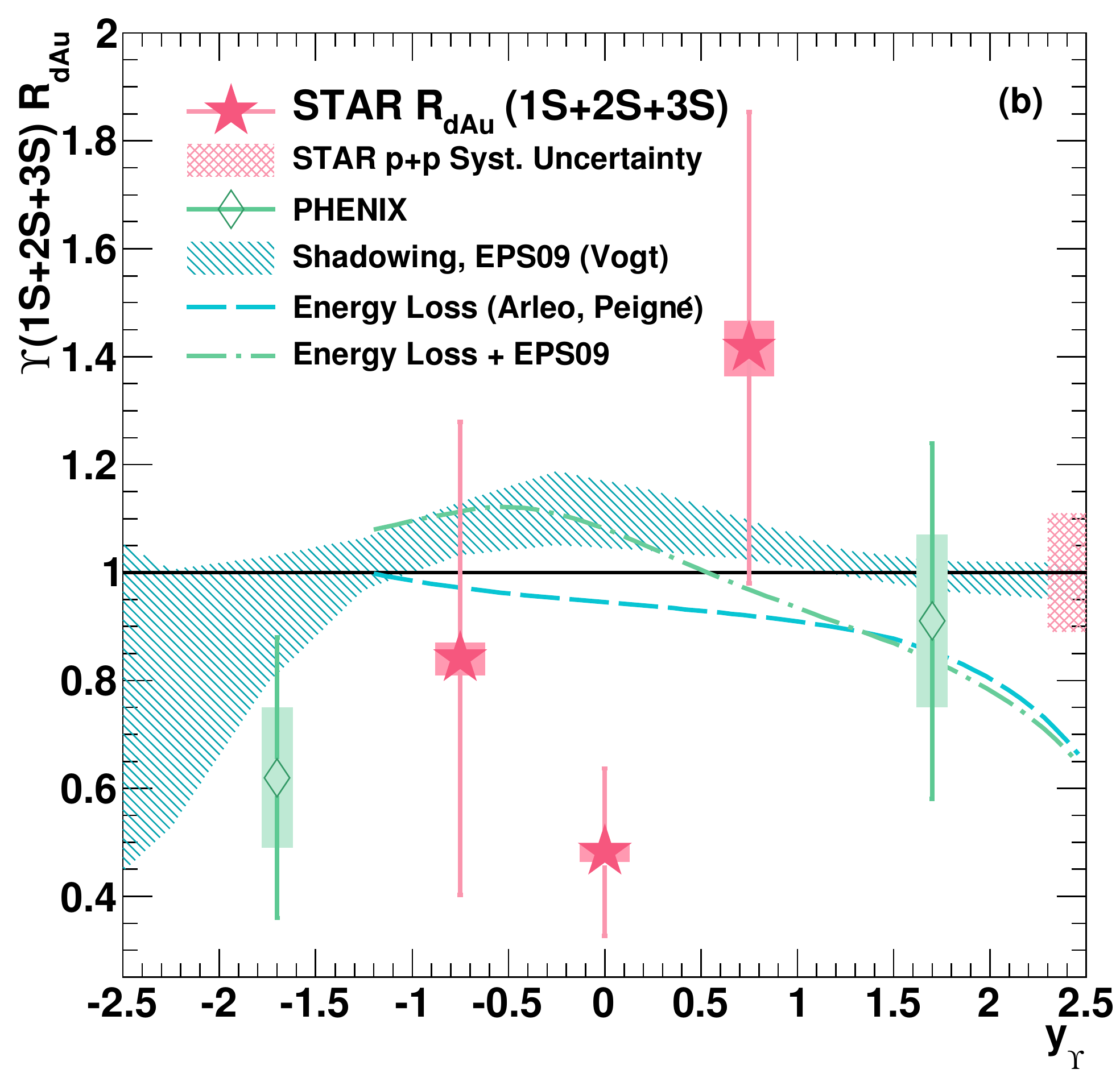}
		\includegraphics[width=0.49\textwidth]{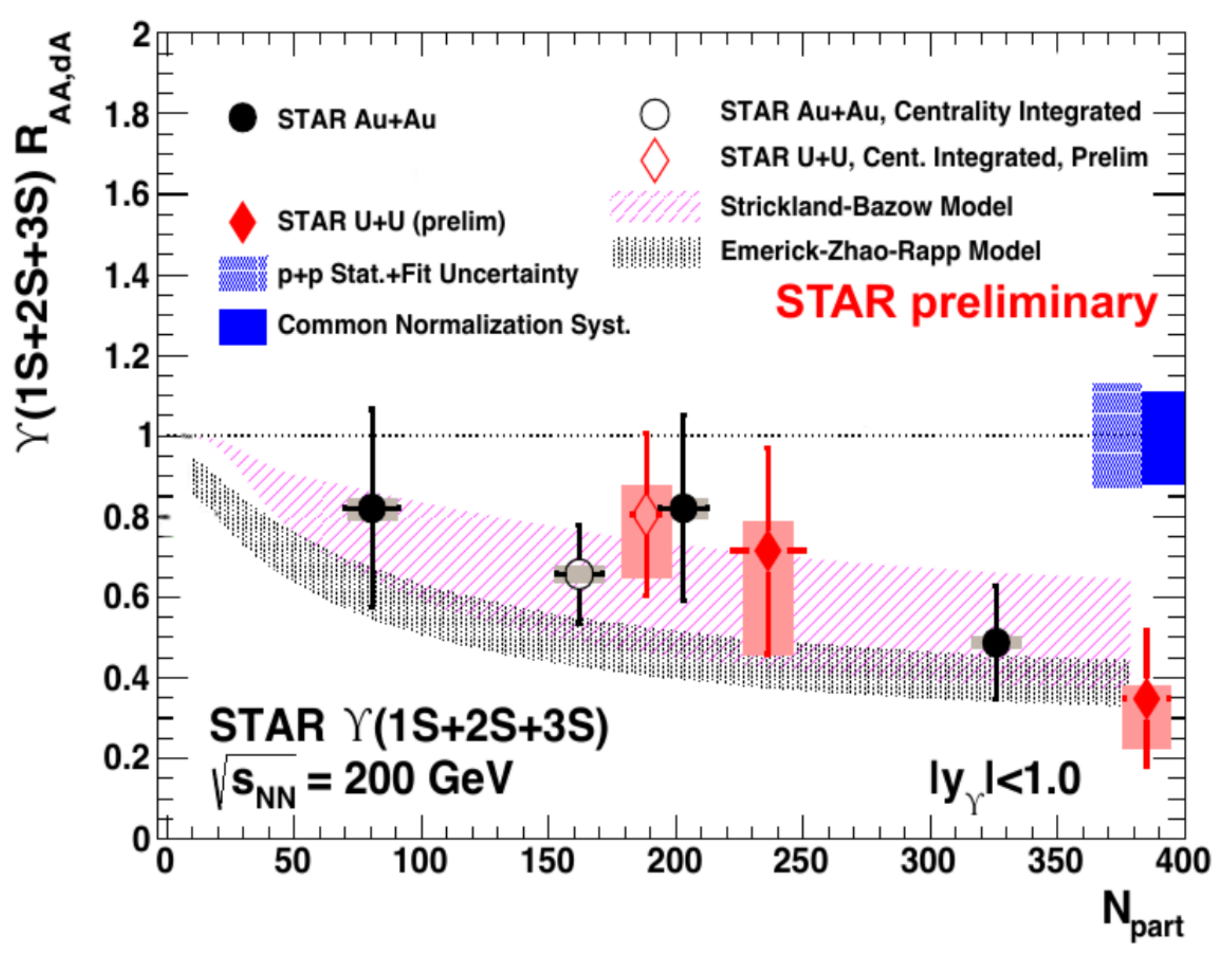}
		\caption{Left: $R_{dAu}$ as a function of rapidity for STAR~\cite{Adamczyk:2013poh} (stars), and PHENIX~\cite{Adare:2012bv} (diamonds) results, compared to different model predictions~\cite{Arleo:2012rs}.
		Right: $R_{AA}$ as a function of $N_{part}$ for $\Upsilon(1S+2S+3S)$ at $\vert y \vert < 1$, Au+Au~\cite{Adamczyk:2013poh} (circles), and U+U (diamonds) collisions, compared to two model predictions (shaded areas)~\cite{Strickland:2011aa, Emerick:2011xu}.}
		\label{fig:upsilon}
\end{figure}

The right panel of \cff{fig:upsilon} presents the nuclear modification factors as a function of $N_{part}$, for $\Upsilon(1S+2S+3S)$ in Au+Au collisions at $\sqrt{s_{NN}} =$ 200 GeV (full circles) and U+U collisions at $\sqrt{s_{NN}} =$ 193 GeV (full diamonds), at $| y | <$ 1.
A strong suppression is observed both in Au+Au: $R_{AuAu} = $0.49$\pm$0.13(Au+Au stat.)$\pm$0.07($p+p$ stat.)$\pm0.02$(Au+Au syst.)$\pm 0.06$($p+p$ syst.) and U+U collisions: $R_{UU} = $0.35$\pm$0.17(stat.)$^{\pm0.03}_{-0.13}$(syst.), in the most central collision. The $R_{AA}$ results in both colliding systems show similar trend. The nuclear modification factor for $\Upsilon(1S)$ ground state shows also a suppression in the most central 10\% Au+Au collisions~\cite{Adamczyk:2013poh} that is in agreement with the prediction of the Liu {\it{et al.}} model~\cite{Liu:2010ej} where the suppression is mostly due to the dissociation of the excited states.
The data are consistent with model predictions of Strickland and Bozov~\cite{Strickland:2011aa}, and Emerick~{\it{et al.}}~\cite{Emerick:2011xu}, that include hot-nuclear-matter effects. Calculations of Emerick {\it{et al.}} in addition take into account the CNM effects.
STAR has also observed the full suppression of $\Upsilon(2S+3S)$ states, the 95\%-confidence upper limit for $R_{AA}$ in the centrality range of 0-60\% is $R_{AA}(2S+3S) <$ 0.32.
However, understanding of CNM effects is important for a better interpretation of results from heavy-ion collisions.

\section{Summary}

In this proceedings we present STAR latest results of the J/$\psi$, $\psi(2S)$ and $\Upsilon$ production in $p+p$, $d+$Au, Au+Au and U+U collisions at different colliding energies.
J/$\psi$ and $\Upsilon$ suppression in central heavy-ion collisions and the indication of the complete suppression of $\Upsilon(2S+3S)$  suggest formation of the Quark-Gluon Plasma.
However, no strong dependence on the colliding energy or colliding system of the suppression is observed.

\section*{Acknowledgements}

This publication was supported by the European social fund within the framework of realizing the project ,,Support of inter-sectoral mobility and quality enhancement of research teams at Czech Technical University in Prague'', \\ CZ.1.07/2.3.00/30.0034 and by Grant Agency of the Czech Republic, grant No.13-20841S. 

\begin
{thebibliography}{999}
\bibitem{Matsui:1986dk}
  T.~Matsui and H.~Satz,
  Phys.\ Lett.\ B {\bf 178} (1986) 416.

\bibitem{Mocsy:2007jz}
  A.~Mocsy and P.~Petreczky,
  Phys.\ Rev.\ Lett.\  {\bf 99} (2007) 211602

\bibitem{Zhao:2010nk}
  X.~Zhao and R.~Rapp,
  Phys.\ Rev.\ C {\bf 82} (2010) 064905

\bibitem{Rapp:2008tf}
  R.~Rapp, D.~Blaschke and P.~Crochet,
  Prog.\ Part.\ Nucl.\ Phys.\  {\bf 65} (2010) 209

\bibitem{Adamczyk:2013poh}
  L.~Adamczyk {\it et al.}  [STAR Collaboration],
  Phys.\ Lett.\ B {\bf 735} (2014) 127
   [Phys.\ Lett.\ B {\bf 743} (2015) 537]

\bibitem{Kosarzewski:2012zz}
  L.~Kosarzewski [STAR Collaboration],
  Acta Phys.\ Polon.\ Supp.\  {\bf 5} (2012) 543.

\bibitem{Adamczyk:2012ey}
  L.~Adamczyk {\it et al.}  [STAR Collaboration],
  Phys.\ Lett.\ B {\bf 722} (2013) 55

\bibitem{Frawley:2008kk}
  A.~D.~Frawley, T.~Ullrich and R.~Vogt,
  Phys.\ Rept.\  {\bf 462} (2008) 125

\bibitem{Ma:2010jj}
  Y.~Q.~Ma, K.~Wang and K.~T.~Chao,
  Phys.\ Rev.\ D {\bf 84} (2011) 114001

\bibitem{Artoisenet:2007qm}
  P.~Artoisenet, F.~Maltoni and T.~Stelzer,
  JHEP {\bf 0802} (2008) 102

\bibitem{Adamczyk:2013vjy}
  L.~Adamczyk {\it et al.}  [STAR Collaboration],
  Phys.\ Lett.\ B {\bf 739} (2014) 180
  
\bibitem{Chung:2009xr}
  H.~S.~Chung, C.~Yu, S.~Kim and J.~Lee,
  Phys.\ Rev.\ D {\bf 81} (2010) 014020
  
\bibitem{Lansberg:2010vq}
  J.~P.~Lansberg,
  Phys.\ Lett.\ B {\bf 695} (2011) 149

\bibitem{Abelev:2009qaa}
  B.~I.~Abelev {\it et al.}  [STAR Collaboration],
  Phys.\ Rev.\ C {\bf 80} (2009) 041902

\bibitem{Artoisenet:2008fc}
  P.~Artoisenet, J.~M.~Campbell, J.~P.~Lansberg, F.~Maltoni and F.~Tramontano,
  Phys.\ Rev.\ Lett.\  {\bf 101} (2008) 152001

\bibitem{Adamczyk:2013tvk}
  L.~Adamczyk {\it et al.}  [STAR Collaboration],
  Phys.\ Rev.\ C {\bf 90} (2014) 2,  024906

\bibitem{Liu:2009nb}
  Y.~p.~Liu, Z.~Qu, N.~Xu and P.~f.~Zhuang,
  Phys.\ Lett.\ B {\bf 678} (2009) 72
  
\bibitem{Nelson:2012bc}
  R.~E.~Nelson, R.~Vogt and A.~D.~Frawley,
  Phys.\ Rev.\ C {\bf 87} (2013) 1,  014908

\bibitem{Kikola:2011zz}
  D.~Kikola, G.~Odyniec and R.~Vogt,
  Phys.\ Rev.\ C {\bf 84} (2011) 054907

\bibitem{Arleo:2012rs}
  F.~Arleo and S.~Peigne,
  JHEP {\bf 1303} (2013) 122

\bibitem{Adare:2012bv}
  A.~Adare {\it et al.}  [PHENIX Collaboration],
  Phys.\ Rev.\ C {\bf 87} (2013) 044909

\bibitem{Strickland:2011aa}
  M.~Strickland and D.~Bazow,
  Nucl.\ Phys.\ A {\bf 879} (2012) 25

\bibitem{Emerick:2011xu}
  A.~Emerick, X.~Zhao and R.~Rapp,
  Eur.\ Phys.\ J.\ A {\bf 48} (2012) 72

\bibitem{Liu:2010ej}
  Y.~Liu, B.~Chen, N.~Xu and P.~Zhuang,
  Phys.\ Lett.\ B {\bf 697} (2011) 32

\end{thebibliography}


\end{document}